\documentstyle[aps,twocolumn,psfig,floats,prc]{revtex}

\begin{document}

\draft

\title{Resilient Reducibility in Nuclear Multifragmentation}

\author{L.G. Moretto, L. Phair, and G.J. Wozniak}

\address{Nuclear Science Division, Lawrence Berkeley National
Laboratory, Berkeley, CA 94720, USA\\}

\date{\today}

\maketitle

\begin{abstract}
The resilience to averaging over an initial energy distribution of
reducibility and thermal scaling observed in nuclear
multifragmentation is studied.  Poissonian reducibility and the
associated thermal scaling of the mean are shown to be
robust. Binomial reducibility and thermal scaling of the elementary
probability are robust under a broad range of conditions. The
experimental data do not show any indication of deviation due to
averaging.

\end{abstract}


The complexity of nuclear multifragmentation underwent a remarkable
simplification when it was {\em empirically} observed that many
aspects of this process were: a) ``reducible''; and b) ``thermally
scalable''
\cite{Moretto95,Tso95,Moretto97,Beaulieu98,Phair95,Phair96}.

``Reducibility'' means that a given many-fragment probability can be
expressed in terms of a corresponding one-fragment probability, i.e.,
the fragments are emitted essentially independent of one another.

``Thermal scaling'' means that the one-fragment probability so
extracted has a thermal-like dependence, i.e., it is essentially a
Boltzmann factor.

Both
``reducibility'' and ``thermal scaling'' were observed in terms of a
global variable, the transverse energy $E_t$ (defined as $E_t=\sum _i
E_i\sin ^2\theta _i$, i.e.~the sum of the kinetic energies $E$ of all
charged particles in an event weighted by the sine squared of their
polar angles $\theta$), which was assumed (see below) to be
proportional to the excitation energy of the decaying source(s)
\cite{Moretto95,Tso95,Moretto97}.

In particular, it was found that the $Z$-integrated multiplicity
distributions $P(n)$ were binomially distributed, and thus
``reducible'' to a one-fragment probability $p$. With higher
resolution, it was noticed that for each individual fragment species
of a given $Z$, the $n_Z$-fragment multiplicities $P(n_Z)$ obeyed a
nearly Poisson distribution, and were thus ``reducible'' to a
single-fragment probability proportional to the mean value $\left<
n_Z\right>$ for each $Z$ \cite{Beaulieu98}.

The one-fragment probabilities $p$ showed ``thermal scaling'' by
giving linear Arrhenius plots of $\ln p$ vs $1/\sqrt{E_t}$ where it is
assumed that $\sqrt{E_t}\propto T$. Similarly $n$-fragment charge
distributions $P_n(Z)$ were shown to be both ``reducible'' to a
one-fragment $Z$ distribution as well as ``thermally scalable''
\cite{Phair95}.  Even the two-fragment angular correlations
$P_{1,2}(\Delta\phi )$ were shown to be expressible in terms of a
one-body angular distribution with amplitudes that are ``thermally
scalable'' \cite{Phair96}.  Table \ref{table:one} gives a summary of
the ``reducible'' and ``thermal scaling'' observables.

{\em Empirically}, ``reducibility'' and ``thermal scaling'' are
pervasive features of nuclear multifragmentation.  ``Reducibility''
proves nearly stochastic emission. ``Thermal scaling'' gives an
indication of thermalization.

Recently, there have been some questions on the significance (not the
factuality) of ``reducibility'' and ``thermal scaling'' in the {\em
binomial} decomposition of $Z$-integrated multiplicities \cite{Tok97}.
For instance, had the original distribution in the true
excitation-energy variable been binomially distributed and thermally
scalable, wouldn't the process of transforming from excitation energy
$E$ to transverse energy $E_t$ through an (assumedly) broad
transformation function $P(E,E_t)$ destroy both features?

\begin{table}
\begin{tabular}{|l|l|l|}
\multicolumn{1}{|l|}{reducibility} &
\multicolumn{1}{l|}{thermal scaling} &
\multicolumn{1}{l|}{reference} \\ \hline
$P(n)\rightarrow p$ & $\ln p 
\propto 1/\protect\sqrt{E_t}$ & \cite{Moretto95} 
\\ 
$P(n_Z)\rightarrow \left< n_Z\right>$ & 
$\ln\left< n_Z\right>\propto 1/\protect\sqrt{E_t}$ & 
\cite{Beaulieu98} 
\\ 
$P_n(Z)\rightarrow P_1(Z)\propto e^{-\alpha Z}$ & $\alpha\propto 
1/\protect\sqrt{E_t}$ & 
\cite{Phair95} 
\\
$P_{1,2}(\Delta\phi)\rightarrow \int P_1(\phi)
P_2(\phi +\Delta\phi) $ &
$ {\rm amplitude}\propto 1/{E_t}$ & 
\cite{Phair96} 
\\
\end{tabular}
\vspace{0.3cm}
\caption{Summary of reducible and thermal 
scaling observables in nuclear 
multifragmentation.}
\label{table:one}
\end{table}
 
Specifically, under a special choice of averaging function (Gaussian),
for a special choice of parameters (variance from GEMINI
\cite{Cha88}), and for special input $p$ (the excitation energy
dependent one-fragment emission probability) and $m$ (the number of
``throws'' or attempts) to the binomial function, the binomial
parameters {\em extracted} from the averaged binomial distribution are
catastrophically altered, and the initial thermal scaling is spoiled
\cite{Tok97}. This ``spoiling'' in \cite{Tok97} is not due to detector
acceptance effects (which has been commented on extensively in
\cite{Moretto97}), but rather is due to the intrinsic width of
correlation between $E_t$ and $E$ as discussed below.

It should be pointed out
that, while the decomposition of the many-fragment emission
probabilities $P(n)$ into $p$ and $m$ may be sensitive to the
averaging process, the quantity $\left< mp\right>$ is not
\cite{Tok97}.  However, both $p$ and $\left< mp\right>$ are known to
give linear Arrhenius plots with essentially the same slope (see
below). This by itself demonstrates that no damaging average is
occurring.

Furthermore, we have observed that by restricting the definition of
``fragment'' to a single $Z$, the multiplicity distributions become
nearly Poissonian and thus are characterized
by the average multiplicity $\left< mp\right>$ which gives well
behaved Arrhenius plots \cite{Beaulieu98}.  {\em Thus, the linearity
of the Arrhenius plots of both $p$ and $\left< mp\right>$ extracted
from all fragments, and the linearity of the Arrhenius plots of
$\left< mp\right>$ for each individual $Z$ value eliminate
observationally the criticisms described above. In fact, it follows
that no visible damage is inflicted by the true physical
transformation from $E$ to $E_t$.
Therefore, the experimental Poisson ``reducibility'' of multiplicity
distributions for each individual $Z$ and the associated ``thermal
scaling'' of the means eliminates observationally these criticisms.}

We proceed now to show in detail that: 1) binomial reducibility and
thermal scaling are also quite robust under reasonable averaging
conditions; 2) the data do not show any indication of pathological
behavior.

We first discuss the possible origin and widths of the averaging
distribution.

It is not apparent why the variance of $P(E,E_t)$ calculated from
GEMINI \cite{Cha88} should be relevant. GEMINI is a low energy
statistical code and is singularly unable to reproduce intermediate
mass fragment (IMF:$3\le Z\le 20$) multiplicities, the magnitudes of
$E_t$, and other multifragmentation features. There is no reason to
expect that the variance in question is realistic.


Apparently, $E_t$
does not originate in the late thermal phase of the reaction.  Rather,
it seems to be dominated by the initial stages of the collision.
Consequently its magnitude may reflect the geometry of the reaction
and the consequent energy deposition in terms of the number of primary
nucleon-nucleon collisions. This is attested to by the magnitude of
$E_t$ which is several times larger than predicted by any thermal
model. Thus, the worrisome ``thermal widths'' are presumably
irrelevant.


Since there is no reliable way to determine the actual resolution of
the correlation between $E_t$ and $E$, experimentally or via
simulation calculations \cite{Tok97}, instead of using large or small
variances, we will show:

a) which variables control the divergence assuming a Gaussian
distribution, and in what range of values the averaging is ``safe'',
i.e.~it does not produce divergent behavior;

b) that the use of Gaussian tails is dangerous and improper unless one
shows that the physics itself {\em requires} such tails.

The input binomial distribution is characterized by $m$, the number of
throws (assumed constant in the calculations in \cite{Tok97}), and $p$
which has a characteristic energy dependence of
\begin{equation}
\log \frac{1}{p}\propto\frac{B}{\sqrt{E_t'}}.
\end{equation}
We denote the one-to-one image of $E$ in $E_t$ space with a prime
symbol.

The averaging in \cite{Tok97} is performed by
integrating the product of an exponential folded with a Gaussian
(Eq.~(12) of \cite{Tok97}).
\begin{equation}
\left< p\right>\propto\int{ \exp\left(
-\frac{B}{\sqrt{x}}-\frac{(x-x_0)^2}{2\sigma ^2} \right) dx }.
\end{equation}
If the slope of the exponential is large, there will be 1) a
substantial shift $\epsilon$ in the peak of the integrand, and 2) a
great sensitivity to the tail of the Gaussian.

The shifts $\epsilon _{\left< p\right>}$ and $\epsilon _{\left<
p^2\right>}$ can be approximately evaluated:
\begin{equation}
\epsilon_{\left< p\right>}=\frac{\sigma ^2B}{2x_0^{3/2}}
\label{eq:shift1}
\end{equation}

\begin{equation}
\epsilon _{\left< p^2\right>} = 2 \frac{\sigma ^2B}{2x_0^{3/2}}.
\label{eq:shift2}
\end{equation}
This illustrates the divergence at small values of $x_0$ both in the
shift of the integrand in $\left< p\right>$ and $\left< p^2\right>$
and the corresponding divergence in $\sigma ^2_p=\left<
p^2\right>-\left< p\right>^2$.  The scale of the divergence is set by
the product $\sigma ^2B$. Thus one can force a catastrophic blowup by
choosing a large value of $\sigma ^2$, of $B$, or of both. This is
what has been shown to happen with
large values of $\sigma ^2$ and $B$. The counterpart to this is that
there possibly exists a range of values for $B$ and $\sigma ^2$ which
leads to a ``safe'' averaging process.

\begin{figure}
\centerline{\psfig{file=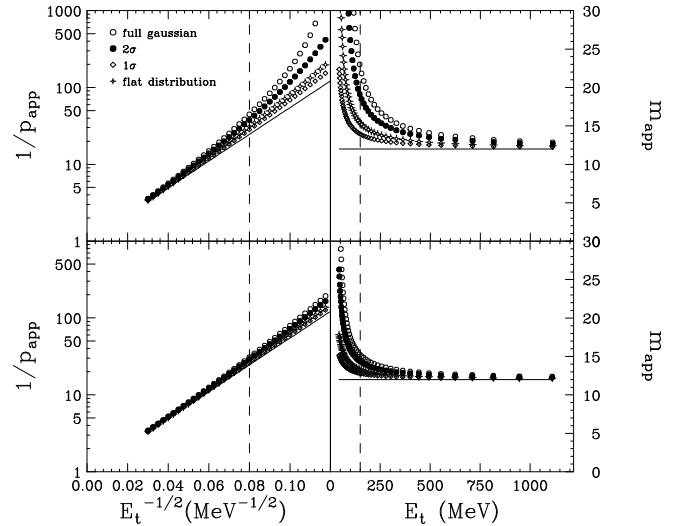,angle=90,height=6.9cm}}
\vspace{0.3cm}
\caption{Upper left panel: 
the distorted Arrhenius plot 
for a value of $B$=40MeV$^{1/2}$, $m$=12, and 
a fixed ratio of $\Gamma_{E_t}/E_t$=0.3. 
The open circles represent the apparent single 
fragment emission probabilities
extracted after a folding of the ``thermal'' 
binomial emission probabilities
with a Gaussian distribution 
(see Eqs.~(\protect\ref{eq:p_app})-(\protect\ref{eq:var_n})). 
The solid circles
and open diamonds show the effect of truncating the Gaussian tails 
at 2$\sigma$ and
1$\sigma$, respectively. The star symbols demonstrate results from 
folding with a square
distribution 
of the same full width at half maximum as the Gaussian.
The solid line represents the
undistorted Arrhenius plot. The dashed line 
represents an experimental ``soft''
limit beyond which the transverse energy may be 
unreliable as a measure of 
impact 
parameter \protect\cite{Phair92b} or
deposited energy. Lower left panel: same as upper left but for 
$\Gamma_{E_t}/E_t$=0.2. Upper right panel: 
the distorted values of the number of throws $m_{app}$ 
as a function of $E_t$ for a value of $B$=40MeV$^{1/2}$ and 
a fixed ratio of $\Gamma_{E_t}/E_t$=0.3. The solid 
line represents the
undistorted value of $m$=12. Lower right panel: same 
as upper right but for 
$\Gamma_{E_t}/E_t$=0.2.}
\vspace{2mm}
\label{fig:composite_pm}
\end{figure}

In order to illustrate this, we have calculated the ``apparent''
values of the single fragment emission probability $p_{app}$ for
widths characterized by the ratio of the full width at half maximum
$\Gamma _{E_t}$ over $E_t$. Specifically we have extracted $p_{app}$:
\begin{equation}
p_{app}=1-\frac{\sigma _n^2}{\left< n\right>}
\label{eq:p_app}
\end{equation}
and $m_{app}$:
\begin{equation}
m_{app}=\frac{\left< n\right>}{p_{app}}
\end{equation}
by calculating the observed mean:
\begin{equation}
\left< n\right>=\int{\sum _{n=0}^m nP_n^m(E_t')g(E_t')dE_t'}
\end{equation}
and variance:
\begin{equation}
\sigma _n^2=\left[ \int{\sum _{n=0}^m
n^2P_n^m(E_t')g(E_t')dE_t'}\right]-\left< n\right>^2
\label{eq:var_n}
\end{equation}
for ``thermal'' emission probabilities $P_n^m$ folded with a Gaussian
distribution $g(E_t)$.  We have assumed $m$ is constant.

For a value of $\Gamma _{E_t}/E_t$=0.3, $m$=12, and $B$=40MeV$^{1/2}$
(consistent with the upper limits of the slopes observed in the Xe
induced reactions \cite{Tso95,Moretto97}), the onset of divergence is
observed in the Arrhenius plot at
small values of $E_t$
(top left panel of Fig.~\ref{fig:composite_pm}, open circles).
For $\Gamma _{E_t}/E_t$=0.2 (open circles in bottom left panel of
Fig.~\ref{fig:composite_pm}), the divergent behavior is ``shifted'' to
even lower energies and the resulting Arrhenius plot remains
approximately linear. Therefore, the thermal signature survives.  For
both widths, the linear (thermal) scaling survives {\em in the
physically explored range} of $1/\sqrt{E_t}\le 0.08$ ($E_t\ge$ 150
MeV) shown by the dashed lines in Fig.~\ref{fig:composite_pm}.  As we
shall see below, the effect is weaker
for even lower values of $B$ which are
commonly seen experimentally.

The divergent behavior manifests itself as well in the parameter $m$,
the number of ``throws'' in the binomial description. Values of
$m_{app}$ are plotted (open circles) as a function of $E_t$ in the
right column of Fig.~\ref{fig:composite_pm} for $\Gamma_{E_t}/E_t$=0.3
(top panel) and $\Gamma_{E_t}/E_t$=0.2 (bottom panel).

While the distortions depend mostly on the variance of the energy
distribution, distributions with
similar widths can be associated with very different variances. For
instance, a Lorentzian distribution with finite $\Gamma$ has infinite
variance. Its use would lead to a divergence even for infinitely small
values of $\Gamma$. Thus, even innocent trimmings to the
(non-physical) tails of a Gaussian can produce big differences in the
variance of the distribution and in the ensuing corrections.  We
exemplify this point in two ways.

a) We use a ``square'' distribution with a width equal to the full
width at half maximum of the Gaussian. As can be seen by the star
symbols of Fig.~\ref{fig:composite_pm} this simple exercise
dramatically extends the range over which the average can be performed
safely.

b) We truncate the tails of the Gaussian at 1$\sigma$ (diamonds) and
 2$\sigma$ (solid circles) in Fig.~\ref{fig:composite_pm}. Already the
 cut at $2\sigma$ shows a dramatic improvement over a full
 Gaussian. The 1$\sigma$ cut actually makes things even better than
 the square distribution (as seen in Fig.~\ref{fig:composite_pm}).

\begin{figure}
\centerline{\psfig{file=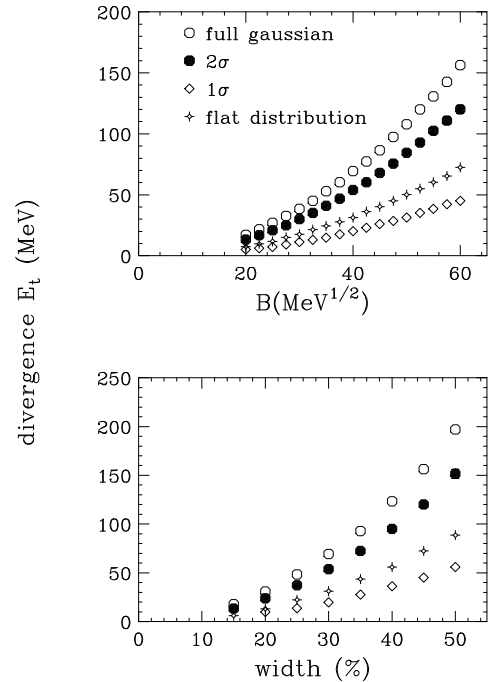,angle=90,height=9cm}}
\vspace{0.3cm}
\caption{Top panel: 
the divergence energy (the energy at which $m_{app}$ and $p_{app}$
change sign) as a function of the slope parameter 
$B$ for a fixed ratio of $\Gamma_{E_t}/E_t$=0.3. Bottom panel:
the divergence energy as a function of the width $\Gamma_{E_t}/E_t$ 
for a fixed value of $B$=40MeV$^{1/2}$. }
\label{fig:B_percent_dep}
\end{figure}

To illustrate the conditions under which the ``thermal'' scaling
survives (i.e.~linear Arrhenius plots as a function of
$1/\sqrt{E_t}$), we have traced the evolution of the ``divergence
energy'' (or the point at which $m_{app}$ and $p_{app}$ change sign)
as a function of the two parameters which control the strength of the
divergence: the slope parameter $B$ and the variance $\sigma ^2$
(hereafter characterized by its full width at half maximum value
$\Gamma _{E_t}\approx 2.35\sqrt{\sigma ^2}$).

A particular example for $\Gamma _{E_t}/E_t$=0.3 is shown by the open
circles in the top panel of Fig.~\ref{fig:B_percent_dep}.  In
addition, values of the divergence energy for 1$\sigma$ and 2$\sigma$
truncations of the Gaussian as well as a square distribution are also
plotted. For all intents and purposes, divergencies that occur at less
than 100 MeV do not alter substantially the linear Arrhenius plots as
they have been observed to date \cite{Moretto95,Tso95,Moretto97} in
the $E_t$ range of 150 to 1600 MeV.

In a similar manner, the dependence of the divergence energy can also
be determined as a function of the relative width $\Gamma_{E_t}/E_t$
(for a fixed value of $B$). This behavior is demonstrated in the
bottom panel of Fig.~\ref{fig:B_percent_dep}.

\begin{figure}
\centerline{\psfig{file=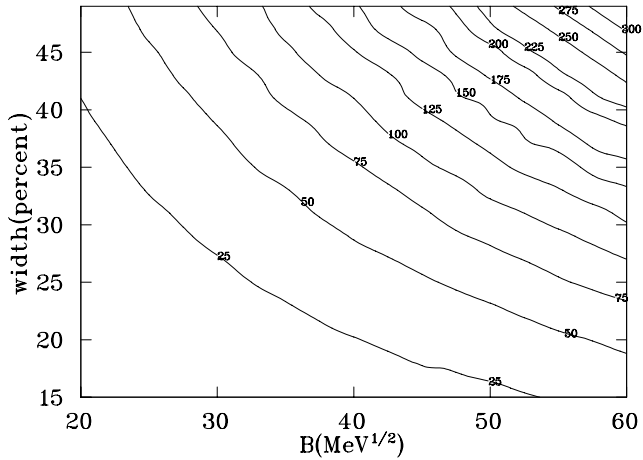,height=6.0cm,angle=90}}
\vspace{0.3cm}
\caption{The divergence energies (contour line) as a 
function of the width
$\Gamma _{E_t}/E_t$ and the slope B for $m$=12 and a 
Gaussian distribution 
truncated at
$2\sigma$.}
\label{div_2d}
\end{figure}

A more global view of the parameter space is shown in
Fig.~\ref{div_2d} where the divergence energy is plotted (contour
lines) as a function of the width $\Gamma _{E_t}/E_t$ and slope
$B$. The shape of the contour lines reflects the $\sigma ^2B$ scale
deduced in Eqs.~(\ref{eq:shift1}) and (\ref{eq:shift2}). The
calculation in ref.~\cite{Tok97} sits nearly in the upper right hand
corner of the graph. But, as is clearly demonstrated, large regions
exist where binomial reducibility and thermal scaling survive (roughly
given by the region with divergence energies less than 100 MeV).

From the above exercises it is concluded that {\em there is abundant
room for the survival of binomiality and thermal scaling}.

In this second part, we show that {\em none} of the symptoms of
divergence are present in the available experimental data
\cite{Moretto95,Tso95,Moretto97}. Furthermore,
the average fragment multiplicity $\left< n\right>$ is expected to be
``distortion free'' \cite{Tok97}. As such, it provides a baseline
reference with which to compare the ``distorted'' variable, $p_{app}$
(to verify whether
the label ``distorted'' is appropriate).  In addition, we can force
the divergence to appear in the data, by artificially broadening the
$E_t$ bins, thus establishing that it is not present with ordinary
(small) $E_t$ bins. Finally, we
show that {\em thermal scaling} is {\em present} and {\em persists} in
the data even when the divergence is forced.

First, we draw attention again to the two pathologic features arising
from excessive averaging. 1) The quantity $m$ diverges near $E_t$=0.~~
2) The quantity $1/p$ suffers a corresponding discontinuity at the
same low energy.

\begin{figure}
\centerline{\psfig{file=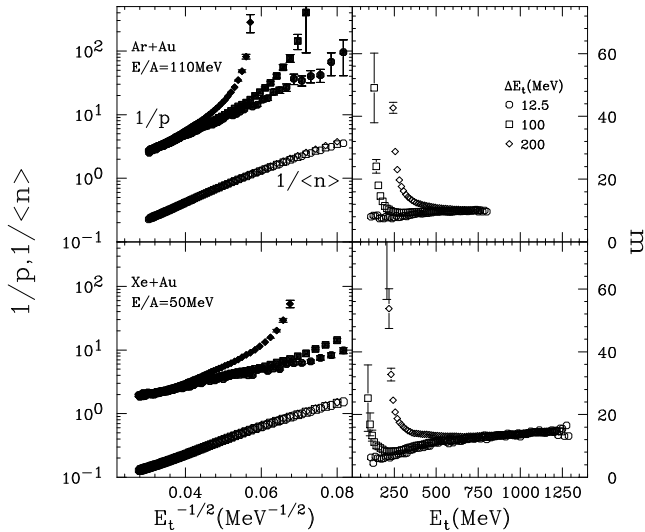,angle=180,height=7.0cm}}
\vspace{0.3cm}
\caption{Left panels: the inverse of the single 
fragment emission probability 
(solid
symbols) and the inverse average fragment 
multiplicity (open symbols) 
as a function of
$1/\protect\sqrt{E_t}$ for the indicated bin 
widths $\Delta E_t$ and systems. 
Right panels: the observed values of $m$ as a 
function of $E_t$ for the 
indicated
bin widths $\Delta E_t$ and systems}
\label{divergence}
\end{figure}

Inspection of the published data shows that:

1) $m$ {\em never diverges near $E_t$ = 0}. To the contrary $m$
remains relatively constant or {\em actually decreases with decreasing
$E_t$}. This is particularly true for all of the Xe induced reactions
\cite{Tso95,Moretto97} (see Fig.~\ref{divergence});

2) $\log 1/p$ is nearly linear vs. $1/\sqrt{E_t}$ over the
experimental $E_t$ range without the indications of trouble suggested
by the calculations in the previous section.

Thus the experimental data do not show any signs of pathological
features.

\begin{figure}
\centerline{\psfig{file=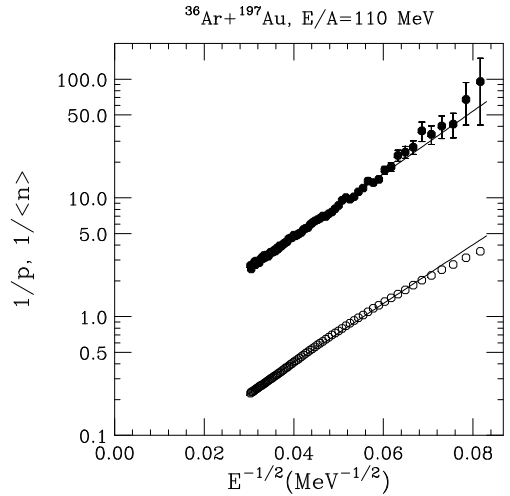,angle=90,height=6.5cm}}
\vspace{0.3cm}
\caption{The inverse of the single fragment emission 
probability (solid
circles) and the inverse of the average fragment 
multiplicity (open circles) 
as a function of
$1/\protect\sqrt{E_t}$ for the reaction Ar+Au at $E/A$=110 MeV.
The solid lines are linear fits to the data.}
\label{inv_p}
\end{figure}

The quantity $\left< n\right>=m_{app}p_{app}$ does not suffer from the
distortions due to averaging.
In fact, $\left< n\right>$ is a suitable alternative for constructing
an Arrhenius plot in those cases where $m$ depends only weakly on
$E_t$ (as observed in many of the data sets we have studied).  A
comparison of the Arrhenius plots constructed from $1/p_{app}$ and
$1/\left< n\right>$ is shown in Fig.~\ref{inv_p}. The striking feature
of this comparison is that the $1/p_{app}$ values {\em have the
same slope} as the ``distortion-free'' case of $1/\left< n\right>$.
Similar observations can be made for all the other reactions studied
so far.  {\em As a consequence both the ``fragile'' $p$ and the
``robust'' $\left< mp\right>$ survive the physical transformation
$P(E,E_t)$ unscathed.}

\begin{figure}
\centerline{\psfig{file=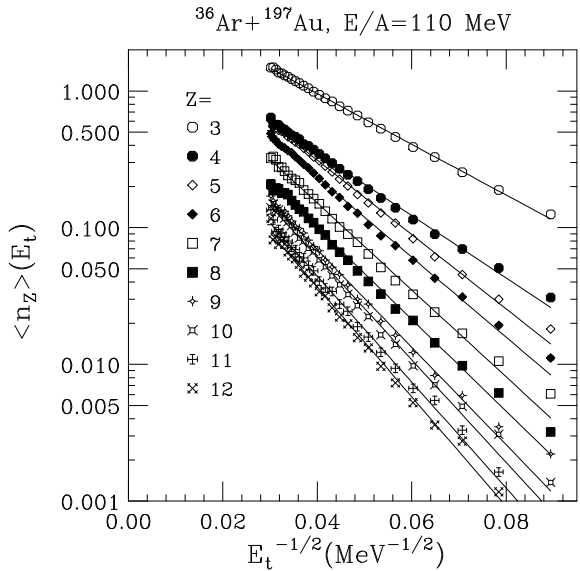,angle=90,height=7.5cm}}
\vspace{0.3cm}
\caption{The average yield per event of elements 
with $Z$ between 3 and 12 
as a function of $1/\protect\sqrt{E_t}$. The
lines are linear fits to the data.}
\label{ave_Z}
\end{figure}

When the probability becomes
small, the binomial distribution reduces to a Poisson
distribution. This can be achieved experimentally by limiting the
selection to a single $Z$ \cite{Beaulieu98}. The observed average
multiplicity is now experimentally equal to the variance. Thus we are
in the Poisson reducibility regime and can check the thermal scaling
directly on $\left< n_Z\right>$.  For a Poisson distribution,
$\log\left< n_Z\right>$ should scale
linearly with $1/\sqrt{E_t}$. This can be seen experimentally for the
average yield of individual elements of a given charge (see
Fig.~\ref{ave_Z}) for the reaction Ar+Au at $E/A$=110 MeV. For the
case of a single species, the reducibility is Poissonian, and the
thermal (linear) scaling with $1/\sqrt{E_t}$ is readily apparent. As
pointed out at the
outset of the paper, this evidence, together with that of
Fig.~\ref{inv_p} {\em indicates that no significant averaging is
occurring even in the case of binomial decomposition.}

The data can
be ``encouraged'' to demonstrate the sort of catastrophic failures
described here.  By widening the bins in transverse energy ($\Delta
E_t$), we can induce an artificial broadening to mimic a broad
correlation between $E$ and $E_t$. For example, the behavior of
$p_{app}$ and $m_{app}$ is shown in Fig.~\ref{divergence}
for three different widths and two different reactions.
The divergencies of $p_{app}$ and $m_{app}$ are readily visible for
large $\Delta E_t$ values, but are noticeably absent for small values.
The spectacularly large binning in $E_t$ (100 MeV!) necessary to force
the anticipated pathologies to appear is reassuring indeed. {\em
Notice that here the absolute width, not the relative width, was kept
fixed even at the lowest energies!}  Furthermore, the stability of
$\left< n\right>$ is readily apparent from the complete overlap of the
values of $\left< n\right>$ extracted for different windows of $E_t$
(open symbols of Fig.~\ref{divergence}).

In summary:

a) Binomial reducibility and the associated thermal scaling survive in
a broad range of parameter space.  The single case shown in
\cite{Tok97} is an extreme one based on unsupported assumptions about
the averaging function.

b) The experimentally observed simultaneous survival of the linear
Arrhenius plot for parameter $p$ and the robust average $\left<
mp\right>$ suggests that no serious damage is generated by the
physical transformation $P(E,E_t)$.

c) The multiplicity distributions for any given $Z$ value are
Poissonian and the resulting average multiplicity $\left<
n\right>=\left< mp\right>$ gives linear Arrhenius plots confirming the
conclusion in b).

d) Finally, the data themselves do not show any indication of
pathological behavior.  This can be seen, for instance, by comparing
the behavior of $p$ with $\left< n\right>$.  The pathology can be
forced upon the data by excessively widening the $E_t$ bins. Even
then, the thermal scaling survives in the average multiplicity.



Acknowledgments

This work was supported by the Director, Office of Energy Research,
Office of High Energy and Nuclear Physics,
Nuclear Physics Division of the US Department of Energy,
under contract DE-AC03-76SF00098.



\end{document}